\documentclass{ifacconf}
\usepackage{color}
\usepackage{tikz}
\usepackage{graphicx}      
\usepackage{natbib}        
\usepackage{amsmath}
\usepackage{xcolor}
\usepackage{amssymb}
\usepackage{commath}
\usepackage{enumerate}
\usepackage{caption}
\usepackage[hyphens]{url}

\definecolor{red_def}{HTML}{d62728}
\newcommand{\redline}{\raisebox{2pt}{\tikz{\draw[dashdotted,red,line width=1](0,0) -- (3.7mm,0);}}}

\graphicspath{{figures/}}
\begin{document}
\begin{frontmatter}

\title{Optimal Synthesis of LTI Koopman Models for Nonlinear Systems with Inputs\thanksref{footnoteinfo}} 

\thanks[footnoteinfo]{This work has received funding from the European Research Council (ERC) under the European Union’s Horizon 2020 research and innovation programme (grant agreement nr. 714663) and from the European Union within the framework of the National Laboratory for Autonomous Systems (RRF-2.3.1-21-2022-00002).}

\author[First]{Lucian C. Iacob} 
\author[First,Second]{Roland Tóth} 
\author[First]{Maarten Schoukens}

\address[First]{Control Systems Group, Eindhoven University of Technology, Eindhoven, The Netherlands \\e-mail: l.c.iacob@tue.nl, r.toth@tue.nl, m.schoukens@tue.nl}
\address[Second]{Systems and Control Laboratory, Institute for Computer Science and Control, Budapest, Hungary}

\begin{abstract}                
A popular technique used to obtain linear representations of nonlinear systems is the so-called Koopman approach, where the nonlinear dynamics are lifted to a (possibly infinite dimensional) linear space through nonlinear functions called observables. In the lifted space, the dynamics are linear and represented by a so-called Koopman operator. While the Koopman theory was originally introduced for autonomous systems, it has been widely used to derive \textit{linear time-invariant} (LTI) models for nonlinear systems with inputs through various approximation schemes such as the \textit{extended dynamics mode decomposition} (EDMD). However, recent extensions of the Koopman theory show that the lifting process for such systems results in a \textit{linear parameter-varying} (LPV) model instead of an LTI form. 
As LTI Koopman model based control has been successfully used in practice and it is generally temping to use such LTI descriptions of nonlinear systems, due to the simplicity of the associated control tool chain, a systematic approach is needed to synthesise optimal LTI approximations of LPV Koopman models compared to the ad-hoc schemes such as EDMD, which is based on least-squares regression.  
 In this work, we introduce optimal LTI Koopman approximations of exact Koopman models of nonlinear systems with inputs by using $\ell_2$-gain and generalized $H_2$ norm performance measures. We demonstrate the advantages of the proposed Koopman modelling procedure  compared to EDMD. \vspace{-2mm}
\end{abstract}


\end{frontmatter}

\section{Introduction} \vspace{-2mm}
{In recent years, significant research has been carried out to embed nonlinear dynamics into linear representations to generalize powerful approaches of the linear framework for analysis and control of systems with dominant nonlinear behavior. One such framework is based on the Koopman operator \citep{Mauroy:20}. In the Koopman approach, so-called observable functions are used to lift the nonlinear state-space to a linear, but possibly infinite dimensional,  representation. While the framework was originally introduced for autonomous systems, recent developments have been made for systems with inputs \citep{Kaiser:21}, \citep{Surana:16}, \citep{Iacob:22}. It has been shown in works such as \citep{Kaiser:21}, \citep{Surana:16} that, in continuous time, the lifted input matrix has a dependency on the state and, in \citep{Iacob:22}, the same property has been shown to hold in discrete time. As  discussed in \citep{Kaiser:21}, \citep{Iacob:22}, the lifted representations can be interpreted as \textit{linear parameter-varying} (LPV) models. While control tools have been developed for LPV systems (see e.g. \citep{Mohammadpour:12}), the use of a purely \textit{linear time-invariant} (LTI) representation is still appealing, due to the simplicity of LTI control methods, like \emph{optimal gain control} and \emph{model predictive control} (MPC), compared to their LPV counterparts. However, existence of purely LTI Koopman representations is only assumed in practice when this concept is applied for systems with input, without considering the  introduced approximation error or trying to systematically mitigate it.

\par The main contribution of this paper is to address this problem by deriving optimal approximations of the input matrix (in an $\ell_2$-gain and generalized $H_2$ sense), starting from the exact LPV Koopman description derived in \citep{Iacob:22} for discrete-time systems. Furthermore, based on \citep{Iacob:22}, we also derive a useful amplitude bound of the state-evolution error that can be used to further compare various LTI approximations in the Koopman setting. We compare the derived methods with the celebrated  \textit{extended dynamics mode decomposition} (EDMD) approach of the Koopman literature \citep{Williams:16, Korda:18}. 
Using a simulation study, we show how much better the state-trajectories associated with the original nonlinear system are represented by the $\ell_2$-gain and $H_2$-norm based synthesis approaches compared to an EDMD-like approximation. 

 The paper is structured as follows. In Section \ref{sec:koop_lift}, the Koopman embedding approach is discussed and the lifted form for discrete-time nonlinear systems with inputs is presented. Section \ref{sec:lti_synthesis} details the proposed synthesis method for optimal approximation of the input matrix. Next, in Section \ref{sec:example}, the approximation error of the LTI models obtained via the introduced synthesis methods is analyzed and compared in a simulation study. Section \ref{sec:conclusion} presents the conclusions.}
\vspace{-.3cm}
\subsubsection{Notation:}
 $\|v\|_2$ stands for the Euclidean norm of a real vector $v\in\mathbb{R}^n$.
 $\rho(A)=\max_{r \in \lambda(A)} | r| $ is the spectral radius of a matrix $A\in\mathbb{R}^{n\times n}$ with  eigenvalues $\lambda(A)$,
while $\bar{\sigma}(P)$ is the largest singular value of $P\in\mathbb{R}^{m\times n}$. $\|P\|_{2,2}$ represents the induced 2,2 matrix norm:
\begin{equation}
\|P\|_{2,2}=\sup_{v\in\mathbb{R}^n \setminus 0}\frac{\|Pv\|_2}{\|v\|_2}=\bar{\sigma}(P).
\end{equation}
For a discrete time signal $v:\mathbb{Z}_+ \rightarrow \mathbb{R}^n$, $\|v\|_2=\sqrt{\sum^\infty_{k=0} \|v_k\|_{2}^2}$, where $v_k\in\mathbb{R}^{n}$ denotes the value of $v$ at time $k$ and $\mathbb{Z}_+$ stands for non-negative integers, and $\|v\|_{\infty}=\max_k \|v_k\|_2$.
\vspace{-.2cm}
\section{Koopman lifting}\label{sec:koop_lift} \vspace{-3mm}
This section details the Koopman lifting approach for autonomous and input driven nonlinear systems, explaining why an LPV model is obtained through lifting in the presence of inputs. Furthermore, we briefly detail the popular EDMD method used for Koopman modelling in practice.
\vspace{-2mm}
\subsection{Lifting for autonomous systems} \vspace{-3mm}
Consider the following nonlinear system:
\begin{equation}\label{eq:nl_aut}
x_{k+1}=f(x_k),
\end{equation}
with $x_k\in\mathbb{X}\subseteq\mathbb{R}^{n_\mathrm{x}}$ being the state variable at time moment $k\in\mathbb{Z}_+$ and $\mathbb{X}$ is an  open connected set, while $f:\mathbb{R}^{n_\mathrm{x}}\rightarrow\mathbb{R}^{n_\mathrm{x}}$ is the nonlinear state transition map. Furthermore, we assume $\mathbb{X}$ to be forward invariant under $f(\cdot)$, i.e. $f(\mathbb{X})\subseteq \mathbb{X}, \forall k\in\mathbb{Z}_+$. The effect of the Koopman operator $\mathcal{K}:\mathcal{F}\rightarrow\mathcal{F}$ associated with \eqref{eq:nl_aut} is described as:
\begin{equation}\label{eq:koop_obs_1}
\mathcal{K}\phi=\phi \circ f,
\end{equation}
with $\mathcal{F}$ being a Banach space of so-called observable (or lifting) functions $\phi:\mathbb{X}\rightarrow\mathbb{R}$.
For an arbitrary state $x_k$, based on \eqref{eq:nl_aut} and \eqref{eq:koop_obs_1}, we can write the following expression:
\begin{equation}
\mathcal{K}\phi(x_k)=\phi \circ f(x_k)=\phi(x_{k+1}).
\end{equation}
This corresponds to the idea behind the Koopman framework, where the focus is on expressing the dynamics of observables, instead of the dynamics of \eqref{eq:nl_aut}. Generally, there can be an infinite number of observables, which is unusable in practice. In literature, there are numerous works treating the finite dimensional approximation of the Koopman operator (see \citep{Williams:16, Bevanda:21, Brunton:21}). 
Hence, we assume that there exists a finite dimensional Koopman subspace $\mathcal{F}_{n_\mathrm{f}}\subseteq \mathcal{F}$ that is invariant (the image of $\mathcal{K}$ is in $\mathcal{F}_{n_\mathrm{f}}$). Given that $\mathcal{K}$ is a linear operator \citep{Mauroy:20}, $\mathcal{K}\phi$ can be expressed as a linear combination of the elements of $\mathcal{F}_{n_\mathrm{f}}$. Let $\Phi=[\ \phi_1\  \dots\ \phi_{n_\mathrm{f}}\ ]^\top$ be a basis of $\mathcal{F}_{n_\mathrm{f}}$. Using the results described in \citep{Mauroy:20}, the application of the Koopman operator on a basis $\phi_j$ can be written as:
\begin{equation}
\mathcal{K}\phi_j=\sum^{n_\mathrm{f}}_{i=1}K_{i,j}\phi_i,
\end{equation}
where $K\in\mathbb{R}^{n_\mathrm{f}\times n_\mathrm{f}}$  (matrix representation of the Koopman operator) with elements $K_{i,j}$, $i,j\in\{1,\ldots, n_\mathrm{f} \}$. Let $A=K^\top$, then, an exact finite dimensional lifted representation of \eqref{eq:nl_aut} is given by:
\begin{equation}\label{eq:koop_aut_finite}
\Phi(x_{k+1})=A\Phi(x_k).
\end{equation}
Using \eqref{eq:nl_aut}, \eqref{eq:koop_aut_finite} can be equivalently expressed as:
\begin{equation}\label{eq:koop_f_aut_finite}
\Phi \circ f(x_k)=A\Phi(x_k).
\end{equation}
Hence, the existence condition of a Koopman invariant subspace can be formulated as \citep{Iacob:22}:
\begin{equation} \label{eq:Koopman:condition}
\Phi \circ f \in \text{span}\{\Phi\}.
\end{equation}
In order to obtain the original state $x_k$, an inverse transformation $\Phi^\dagger ( \Phi(x_k))=x_k$ is assumed to exist. This is commonly   accomplished by either considering the states to be in the span of the lifting functions or by explicitly including them as observables (identity basis). This has ensured the practical applicability of Koopman solutions \citep{Korda:18}, \citep{Brunton:21}.\par
Finally, to express the LTI representation of \eqref{eq:nl_aut} through the derived Koopman lifting \eqref{eq:koop_aut_finite}, introduce $z_k=\Phi(x_k)$, which gives the following equation:
\begin{equation}
z_{k+1}=Az_k, \quad \text{with } z_0=\Phi(x_0). 
\end{equation}
\subsection{Lifting for systems with inputs} \vspace{-3mm}
Consider the control affine nonlinear system:
\begin{equation}\label{eq:nl_ca}
x_{k+1}=f(x_k)+g(x_k)u_k,
\end{equation}
where $g:\mathbb{R}^{n_{\mathrm{x}}\times n_{\mathrm{u}}}\rightarrow\mathbb{R}^{n_{\mathrm{x}}}$ and $u_k\in\mathbb{U}\subseteq \mathbb{R}^{n_{\mathrm{x}}}$ with $\mathbb{U}$ being an  open connected set. The analytical derivation of the lifted Koopman model associated to \eqref{eq:nl_ca} is given in \citep{Iacob:22}. Note that the control affine form \eqref{eq:nl_ca} is selected for brevity. The methods developed in this paper can also be applied to general systems of the form $x_{k+1}=f(x_k,u_k)$, by factorizing the resulting Koopman model as described in \citep{Iacob:22}. Next we give the theorem that summarizes the results of the aforementioned paper in terms of the exact lifted representation of \eqref{eq:nl_ca}.
\begin{thm}
Given the nonlinear system \eqref{eq:nl_ca} and a lifting function $\Phi$ of class $\mathcal{C}^1$ (continuously differentiable) such that $\Phi(f(\cdot))\in \text{span}\left\lbrace\Phi\right\rbrace$, with $\Phi:\mathbb{X}\rightarrow\mathbb{R}^{n_\mathrm{f}}$ and $\mathbb{X}$ convex, then there exists an exact finite dimensional lifted form defined as:
\begin{equation}\label{eq:dt_koop_exact_lifted}
\Phi(x_{k+1})=A\Phi(x_k)+\mathcal{B}(x_k,u_k)u_k,
\end{equation}
with $A\in\mathbb{R}^{n_{\mathrm{f}}\times n_{\mathrm{f}}}$ and
\begin{equation}\label{eq:dt_B_koop_general}
\begin{split}
\mathcal{B}(x_k&,u_k) = \\
&\left(\int^1_0\frac{\partial \Phi}{\partial x}(f(x_k)+\lambda g(x_k)u_k)\dif \lambda\right)g(x_k).
\end{split}
\end{equation}
\end{thm}
\begin{pf}
See \citep{Iacob:22}.
\end{pf}
\vspace{-.15cm}
Note that \eqref{eq:dt_koop_exact_lifted} can be expressed as an LPV representation. Let $z_k=\Phi(x_k)$ and $B_\mathrm{z}\circ (\Phi,\text{id})=\mathcal{B}$, where $\text{id}$ is the identity function, i.e. $u=\text{id}(u)$. By introducing a scheduling map $p_k=[\ z^\top_k\ \  u^\top_k\ ]^\top$ with $p_k\in\mathbb{P}=\Phi(\mathbb{X})\times \mathbb{U}$, the LPV Koopman model associated with \eqref{eq:nl_ca} is:
\begin{equation}\label{eq:koop_lpv}
z_{k+1}=Az_k+B_\mathrm{z}(p_k)u_k.
\end{equation}
with $z_0=\Phi(x_0)$.
\vspace{-.2cm}
\subsection{EDMD-based lifting}\vspace{-3mm}
A usual numerical method to obtain Koopman forms from data is EDMD \citep{Williams:16, Bevanda:21}. While it also offers the possibility to inspect the spectral properties of the Koopman operator, it is most commonly used to compute the Koopman matrix $A$ through  least-squares regression using an observed data sequence (or grid points in $\mathbb{X}$) of \eqref{eq:nl_aut} in terms of $\mathcal{D}_N=\{x_k\}_{k=0}^{N}$. Based on a dictionary of a priory chosen observable functions, the approach involves the construction of data matrices $Z=[\ \Phi(x_0)\ \dots\ \Phi(x_{N-1})\ ]$ and $Z^+=[\ \Phi(x_1)\ \dots\ \Phi(x_{N})\ ]$, which are shifted one time step with respect to each other. Common choices for the lifting functions include polynomials, radial basis functions or trigonometric functions. Next, by considering a linear relation between the data matrices:
\begin{equation}  \label{eq:EDMD:aut:a}
Z^+ = AZ+E,
\end{equation}
with $E \in \mathbb{R}^{n_\mathrm{f} \times N}$ being the residual error, the Koopman matrix $A$ via this approach is ``computed'' as:
\begin{equation} \label{eq:EDMD:aut}
A = Z^+Z^\dagger,
\end{equation}
where $\dagger$ is the peseudoinverse. Note that if the dictionary $\Phi$ enables a finite dimensional Koopman representation of the system, i.e. \eqref{eq:Koopman:condition} holds, and $\mathcal{D}_N$ is such that $\mathrm{rank}(Z)=n_\mathrm{f}$, then $E$ becomes zero in terms of \eqref{eq:EDMD:aut} and $A$ is equivalent with the result of the analytical lifting. Otherwise, $A$ corresponds only to the $\ell_2$-optimal solution of \eqref{eq:EDMD:aut:a} under $\mathcal{D}_N$ in terms of minimization of $E$.

In several papers \citep{Korda:18, Proctor:16}, this approach has been extended to systems with inputs and the approximation is based on the assumption of linear lifted dynamics:
\begin{equation}\label{eq:lti_form}
z_{k+1}\approx Az_k+Bu_k,
\end{equation}
giving the linear matrix relation:
\begin{equation}
Z^+ = AZ + BU + E. 
\end{equation}
Similarly to the autonomous case, the state and input matrices of the lifted representation are ``computed'' as follows:\vspace{-.2cm}
\begin{equation}
[A\; B] = Z^+ \begin{bmatrix}
Z\\ U
\end{bmatrix}^\dagger,
\end{equation}
with $U=[\ u_0\ \dots\ u_{N-1}\ ]$. The LTI form \eqref{eq:lti_form} has been extensively used in control related papers such as \citep{Korda:18, Ping:21}. However, the LTI nature of the lifted Koopman model is only assumed, without regard to the original nonlinear system and without an elaborate discussion on the induced approximation error w.r.t. \eqref{eq:nl_ca} or \eqref{eq:koop_lpv}. 
\vspace{-.2cm}
\section{Synthesis of LTI Koopman models}\label{sec:lti_synthesis} \vspace{-2mm}
\begin{figure}[t!]
\begin{center}
\includegraphics[width=.32\textwidth]{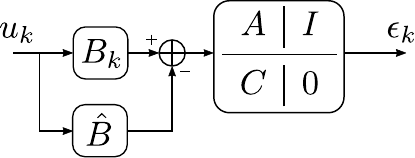}    
\caption{Block diagram of the error system \eqref{eq:err_dynamics}.} 
\label{fig:drawing}
\end{center}
\end{figure}
This section details the synthesis of an approximate constant input matrix $\hat{B}$  to obtain the LTI Koopman model:
\begin{equation}\label{eq:koopman_lti_approx}
\hat{z}_{k+1}=A\hat{z}_k+\hat{B}u_k,
\end{equation}
where $\hat{z}_k \in \mathbb{R}^{n_\mathrm{f}}$ represents the state of this approximate lifted representation at time $k$. The goal is to minimize the approximation error between the LPV Koopman representation \eqref{eq:koop_lpv} and \eqref{eq:koopman_lti_approx}. To give guaranteed bounds, we synthesize the input matrix $\hat{B}$ based on the $\ell_2$-gain and generalized $H_2$ norm performance criteria. At the end of the section, based on \citep{Iacob:22}, we also provide bounds on the induced approximation error under any given input matrix $\hat{B}$. \par The error dynamics between the LPV Koopman model \eqref{eq:koop_lpv} and the approximated LTI system \eqref{eq:koopman_lti_approx} can be written as follows:\vspace{-.1cm}
\begin{subequations}\label{eq:err_dynamics}
\begin{align}
e_{k+1}&=Ae_k+(B_k-\hat{B})u_k, \label{eq:error_in_z}\\
\epsilon_k &= Ce_k,\label{eq:error_in_z_out}
\end{align}
\end{subequations}
with $e_k=z_k-\hat{z}_k$, $B_k=B_\mathrm{z}(p_k)$, while $z_0=\hat{z}_0$, such that $e_0=0$. To obtain the original states, we assume they are in the span of the lifted states, i.e. $x_k=Cz_k$ and $\hat{x}_k=C\hat{z}_k$. Furthermore, $A$, $C$ and $B_k$ are assumed to be known and $A$ satisfies the embedding condition \eqref{eq:koop_f_aut_finite}.\par
To characterise the goodness of the LTI approximation it is a viable approach to consider system norms of \eqref{eq:err_dynamics} such as  $\ell_2$-gain or generalized $H_2$. Analysis and control synthesis based on the $\ell_2$-gain and the generalized $H_2$ norm performance criteria are strongly linked to the notion of dissipativity. As detailed in \citep{Byrnes:94}, a system \eqref{eq:err_dynamics} is dissipative w.r.t. a supply function $s:\mathbb{U}\times \mathbb{X}\rightarrow \mathbb{R}$ if there exists  a positive definite storage function $V:\Phi(\mathbb{X})\rightarrow \mathbb{R}^+$ with $V(0)=0$, such that for all $k\in\mathbb{N}$:
\begin{equation}\label{eq:dissipativity_1}
V(e_{k+1})-V(e_k)\leq s(u_k,\epsilon_k),
\end{equation}
or, equivalently:\vspace{-.1cm}
\begin{equation}
V(e_k)-V(e_0)\leq\sum^{k-1}_{j=0}s(u_j,\epsilon_j),
\end{equation}
where performance measures such as the $\ell_2$-gain or the generalized $H_2$ norm correspond to particular choices of the supply function $s$ \citep{Koelewijn:21, Scherer:15, Verhoek:21}. Note that for the tractability of characterizing the approximation error via dissipativity of 
\eqref{eq:err_dynamics}, the error system is required to be stable, i.e. $\rho(A)<1$. In the following subsections, we derive synthesis approaches to obtain an optimal $\hat{B}$ in the $\ell_2$-gain or generalized $H_2$ norm sense. \vspace{-.2cm}
\subsection{Optimal $\ell_2$-gain approximation}\label{sec:l2_gain}
\vspace{-.2cm}
A commonly used performance measure for LTI systems is the $H_\infty$ norm, which for stable LTI systems corresponds to the $\ell_2$-gain. For the $\ell_2$-gain, the used supply function is $s(u_k,\epsilon_k)=\gamma^2 \|u_k\|^2_2-\|\epsilon_k\|^2_2$. Consider a storage function $V(e_k)=e^\top_kPe_k$ with $P=P^\top\succ 0$. Expanding \eqref{eq:dissipativity_1}:
\begin{multline} \label{eq:l2:cond}
(Ae_k+(B_k-\hat{B})u_k)^\top P(Ae_k+(B_k-\hat{B})u_k)-\\
-e^\top_kPe_k \leq \gamma^2 u^\top_ku_k - \epsilon^\top_k \epsilon^\top_k.
\end{multline}
Next, using  \eqref{eq:error_in_z_out}, the Schur complement and applying a congruence transform with $\text{diag}(P^{-1}, P^{-1}, I_{n_\mathrm{u}}, I_{n_\mathrm{x}})$, we can transform \eqref{eq:l2:cond} to the following set of \textit{linear matrix inequalities} (LMIs):
\begin{equation}\label{eq:l2_lmi}
\begin{bmatrix}
X & AX & B_k - \hat{B} & 0\\ XA^\top & X & 0 & XC^\top \\ B^\top_k - \hat{B}^\top & 0 & \gamma I_{n_\mathrm{u}} & 0 \\ 0 & CX & 0 & \gamma I_{n_\mathrm{x}} 
\end{bmatrix}\succ 0,\quad X=X^\top\succ 0,
\end{equation}
with $X=\gamma P^{-1}$. This set of LMIs needs to hold $\forall k\in\mathbb{Z}_+$ along all possible solution trajectories $(x_k,u_k)$ of \eqref{eq:nl_ca}, i.e., for all $p_k\in\mathbb{P}$. This result resembles \citep{Caigny:13}, with the difference that we do not use a slack variable. The synthesis problem with the decision variables $X$ and $\hat{B}$ can be solved 
in several ways:
\begin{enumerate}[(i)]
\item Introduce a mapping $\mu : \mathbb{P} \rightarrow \mathbb{R}^{n_\delta}$, with minimal $n_\delta>0$ such that  $B_k=B_\mathrm{z}(p_k)={B}_\mathrm{d}(\delta_k)$ and ${B}_\mathrm{d}$ is affine in $\delta_k=\mu(p_k)\in\mu(\mathbb{P})$. Let $\Delta$ be an $n$-vertex polytopic hull of $\mu(\mathbb{P})$. Then, minimize $\gamma$ in $\hat{B}$ such that the LMIs \eqref{eq:l2_lmi} hold at the $n$ vertices of $\Delta$.
\item Consider a mapping $\mu : \mathbb{P} \rightarrow \mathbb{R}^{n_\delta}$ as in (i), but allowing $\hat{B}_\mathrm{z}$ to have  polynomial or rational dependency on $\delta_k$. Then, the minimization of $\gamma$ in $\hat{B}$ can be turned into a convex semi-definite program  by using a \textit{linear fractional representation} (LFR) and a full block S procedure  \citep{Scherer:01}.
\item Another option is to grid the $\mathbb{P}$ space and solve a number of LMIs equal to the number of grid points.
\end{enumerate}
For the sake of simplicity, in the sequel we choose option (iii) to synthesize $\hat{B}$ via gridding and avoid the introduction of any conservativeness via the polytopic embedding in (i) and (ii) at the expense of higher computational load and non-global guarantees. For this, we consider the value of $B_k=B_\mathrm{z}(p_k)$ at grid points $p_k^\mathrm{g}$ obtained as follows. Let  $x^\mathrm{g}_{\cdot,i}\in\mathcal{X}_i\subset \mathbb{X}$ and $u^\mathrm{g}_{\cdot,j}\in\mathcal{U}_j\subset \mathbb{U}$, with $\mathcal{X}_i=\lbrace x^\mathrm{g}_{1,i},\dots,x^\mathrm{g}_{N_i,i}\rbrace$ and $\mathcal{U}_j=\lbrace u^\mathrm{g}_{1,j},\dots,x^\mathrm{g}_{M_j,j}\rbrace$. Here, $x^\mathrm{g}_{k,i}$ denotes the $k^{\text{th}}$ grid value for the $i^{\text{th}}$ state. Similarly, $u^\mathrm{g}_{k,j}$ denotes the $k^{\text{th}}$ grid value of the $j^{\text{th}}$ input channel. Overall, this provides  $\left(\Pi^{n_\mathrm{x}}_{i=1}N_i\right)(\Pi^{n_{\mathrm{u}}}_{j=1}M_j)$ grid points $(x_k^\mathrm{g},u_k^\mathrm{g})$ for which $p_k^\mathrm{g}=[\ \Phi^\top(x_k^\mathrm{g})\ \  u^{\mathrm{g}\top}_k\ ]^\top$ with $\mathcal{P}=\Phi(\mathcal{X})\times \mathcal{U}$. Then, computing $\hat{B}$ amounts to solving the following minimization problem:
\begin{equation*}\label{eq:min_pb_1}
\begin{split}
\min\; & \gamma\\
\text{s.t. }& \begin{bmatrix}
X & AX &  B_\mathrm{z}(p^\mathrm{g}) - \hat{B} & 0\\ XA^\top & X & 0 & XC^\top \\ B_\mathrm{z}^\top(p^\mathrm{g}) - \hat{B}^\top & 0 & \gamma I_{n_\mathrm{u}} & 0 \\ 0 & CX & 0 & \gamma I_{n_\mathrm{x}} 
\end{bmatrix}\succ 0, \\
& \forall p^\mathrm{g}\in\mathcal{P},\quad X=X^\top\succ 0,
\end{split}
\end{equation*}
where the optimum is denoted as $\gamma_{\ell_2}$ with an associated solution $ \hat{B}_{\ell_2}$.
Note that the number of LMIs corresponds to the total number of grid points. The synthesized $\hat{B}_{\ell_2}$ guarantees that the system \eqref{eq:err_dynamics} satisfies the bound:
\begin{equation}
\sup_{0<\|u\|_2<\infty}\frac{\|\epsilon\|_2}{\|u\|_2}<\gamma_{\ell_2}. \vspace{-.2cm}
\end{equation}
\subsection{Generalized $H_2$ norm optimal Koopman model}
\vspace{-.3cm}
Another common performance measure for LTI systems is the generalized $H_2$ norm or 'energy to peak' norm. Similar to the $\ell_2$-gain approach, we use the dissipativity notion to derive the synthesis LMIs to find the optimal $\hat{B}$ in a generalized $H_2$ norm sense. The corresponding supply function is $s(u_k,\epsilon_k)=\gamma\|u_k\|^2_2$.  Under a positive definite storage function, expanding \eqref{eq:dissipativity_1} gives:
\begin{equation}
\begin{split}
(Ae_k+(B_k-\hat{B})u_k)^\top &P(Ae_k+(B_k-\hat{B})u_k)-\\
&-e^\top_kPe_k \leq \gamma u^\top_ku_k.
\end{split}
\end{equation}
Using the Schur complement and a congruence transformation with $\text{diag}(P^{-1},P^{-1},I_{n_\mathrm{u}})$, we obtain the following set of LMIs:
\begin{equation}\label{eq:h2_lmi_1}
\begin{bmatrix}
X & AX & B_k-\hat{B}\\ XA^\top & X & 0 \\ B^\top_k-\hat{B} & 0 & \gamma I_{n_\mathrm{u}}
\end{bmatrix}\succ 0, \quad X=X^\top \succ 0,
\end{equation}
with $X=P^{-1}$, which again needs to hold along all possible solution trajectories $(x_k,u_k)$ of \eqref{eq:nl_ca}, i.e., for all $p_k\in\mathbb{P}$. Furthermore, as detailed in \citep{Scherer:15}, the aim is to satisfy the bound
\begin{equation}
\sup_{0<\|u\|_2<\infty}\frac{\|\epsilon\|_\infty}{\|u\|_2}< \gamma.
\end{equation}
For this, an extra set of LMIs is needed, namely:
\begin{equation}\label{eq:h2_lmi_2}
\begin{bmatrix}
X & XC^\top \\ CX & \gamma I_{n_\mathrm{x}}
\end{bmatrix}\succ 0.
\end{equation}
The validity of this set of LMIs is shown in works such as \citep{Scherer:15, Verhoek:21}, that use an equivalent representation of the matrix inequalities \eqref{eq:h2_lmi_2}. Thus, using the LMIs \eqref{eq:h2_lmi_1} and \eqref{eq:h2_lmi_2}, a synthesis problem can be formulated, that can be solved using the same techniques as mentioned in Section \ref{sec:l2_gain}. Here, we use the previously mentioned gridding approach to solve the following optimization problem:
\begin{equation*}\label{eq:min_pb_2}
\begin{split}
\min\; & \gamma\\
\text{s.t. }& \eqref{eq:h2_lmi_1} \text{ and } \eqref{eq:h2_lmi_2} \text{ with } B_k=B_\mathrm{z}(p^\mathrm{g})  \text{ holds } \forall p^\mathrm{g}\in \mathcal{P},
\end{split}
\end{equation*}
where the optimum is denoted as $\gamma_{H_2}$ with an associated solution $ \hat{B}_{H_2}$.
\vspace{-.3cm}
\subsection{Amplitude bound of the state evolution error}
\vspace{-.3cm}
For a given $\hat{B}$ in \eqref{eq:koopman_lti_approx}, e.g. obtained via EDMD, the previous results can be directly applied to characterise the approximation error in terms of the $\ell_2$-gain or the generalized $H_2$ norm. However, we can also provide an amplitude bound of the state approximation error $e_k$ in \eqref{eq:err_dynamics} based on \citep{Iacob:22}:
\begin{thm}
Consider the LPV Koopman embedding \eqref{eq:koop_lpv} of a general nonlinear system \eqref{eq:nl_ca} and the approximative LTI Koopman form  \eqref{eq:koopman_lti_approx}. Under any initial condition $z_0=\Phi(x_0)=\hat{z}_0$ and input trajectory $u:\mathbb{Z}_+ \rightarrow \mathbb{R}^{n_\mathrm{u}}$ with bounded $\|u\|_\infty$, the error $e_k$ of the state evolution between these representations given by \eqref{eq:error_in_z} satisfies for any time moment $k\in\mathbb{Z}_+$ that:
\begin{enumerate}[(i)]
\item If $\rho(A)<1$, $\|e_k\|_2$ is finite and $\lim_{k\rightarrow\infty}\|e_k\|_2$ exists; \label{enum:item_1_err}
\item If $\bar{\sigma}(A) < 1$, \eqref{enum:item_1_err} is satisfied and furthermore
\begin{equation}\label{eq:err_bound_thm}
\|e_k\|_2\leq \frac{\beta}{1-\bar{\sigma}(A)}\|u\|_{\infty}=\gamma_\text{amp},
\end{equation}
where $\beta=\max_{x\in\mathbb{X},u\in\mathbb{U}} \|B_\mathrm{z}(\Phi(x),u)-\hat{B}\|_{2,2}$.
\end{enumerate}
\end{thm}
\begin{pf}
See \citep{Iacob:22}.
\end{pf}
\vspace{-.3cm}
\section{Numerical example}\label{sec:example}
\vspace{-.3cm}
\subsection{Lifting and simulation}\label{sec:simulation}
\vspace{-.3cm}
\begin{figure}[t!]
\begin{center}
\includegraphics[width=.45\textwidth]{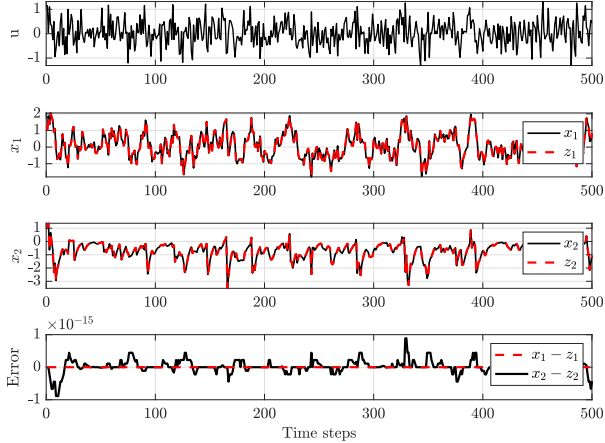}\vspace{-.4cm}    
\caption{Simulation of the state trajectories of the nonlinear system \eqref{eq:nl_dt_ex} in black and exact LPV model \eqref{eq:koop_model_example} given in red, under white noise excitation.} 
\label{fig:simulation}
\end{center}
\end{figure}
Consider the following control affine nonlinear system:
\begin{equation}\label{eq:nl_dt_ex}
x_{k+1} = \underbrace{\begin{bmatrix}
a_1x_{k,1} \\ a_2x_{k,2} - a_3x^2_{k,1}
\end{bmatrix}}_{f(x_k)}+ \underbrace{\begin{bmatrix}
1 \\ x^2_{k,1}
\end{bmatrix}}_{g(x_k)}u_k. \vspace{-.1cm}
\end{equation}
In this example, the notation $x_{k,i}$ denotes the $i^{\text{th}}$ state at time $k$. To simulate the system, we choose the parameters $a_1=a_2=0.7$ and $a_3=0.5$. To lift the system, we select the observables $\Phi^\top (x_k)=[\phi_1(x_k)\; \phi_2(x_k)\; \phi_3(x_k)]=[x_{k,1}\; x_{k,2}\; x^2_{k,1}]$ in order to obtain a finite dimensional lifting of the autonomous part:
\begin{equation}\label{eq:Koopman_matrix_A}
\Phi(x_{k+1})= \underbrace{\begin{bmatrix}
a_1 & 0 & 0 \\ 0 & a_2 & -a_3 \\ 0 & 0 & a^2_1
\end{bmatrix}}_{A}\Phi(x_k). \vspace{-.1cm}
\end{equation}
Based on \eqref{eq:dt_B_koop_general}, the input matrix function is computed as:
\begin{equation}
\label{eq:B_computation_dt}
\begin{split}
\mathcal{B}(x_k,u_k) &=\left(\int^1_0 \begin{bmatrix}
1 & 0 \\ 0 & 1 \\ 2(a_1x_{k,1} + \lambda u_k) & 0
\end{bmatrix}\dif \lambda\right)\begin{bmatrix}
1 \\ x^2_{k,1}
\end{bmatrix}\\&= \begin{bmatrix}
1 \\ x^2_{k,1} \\ 2a_1x_{k,1}+u_k
\end{bmatrix}.
\end{split}
\end{equation}
Thus, the lifted form of \eqref{eq:nl_dt_ex} is:
\begin{equation}
\begin{split}
\Phi(x_{k+1})&=A\Phi(x_k)+\mathcal{B}(x_k,u_k)u_k\\
x_k&=C\Phi(x_k).
\end{split}
\end{equation}
Let $z_k=\Phi(x_k)$. As $x_{k,1}$ and $x^2_{k,1}$ are part of the observable functions, we can define $B_\mathrm{z}\circ (\Phi, \text{id})=
\mathcal{B}$ and  write the LPV Koopman representation of \eqref{eq:nl_dt_ex} as:
\begin{equation}\label{eq:koop_model_example}
\begin{split}
z_{k+1}&=Az_k+B_\mathrm{z}(p_k)u_k\\
x_k&=Cz_k,
\end{split}
\end{equation}
with $p_k = [z_k^\top \; u_k]^\top$ and $C=[I_2 \; 0_{2\times 1}]$. The initial conditions are considered to be $x_0=[1\; 1]^\top$ and $z_0=[1\;1\;1]^\top$. Fig. \ref{fig:simulation} shows the simulated trajectories of the nonlinear system \eqref{eq:nl_dt_ex} and the LPV Koopman representation \eqref{eq:koop_model_example} under white noise excitation $u_k\sim\mathcal{N}(0,0.5)$. It can be observed that the error between the states computed via \eqref{eq:nl_dt_ex} and \eqref{eq:koop_model_example} is negligible and close to numerical precision.
\subsection{LTI synthesis}
\vspace{-.3cm}
\begin{figure}[t!]
\begin{center}
\includegraphics[width=.45\textwidth]{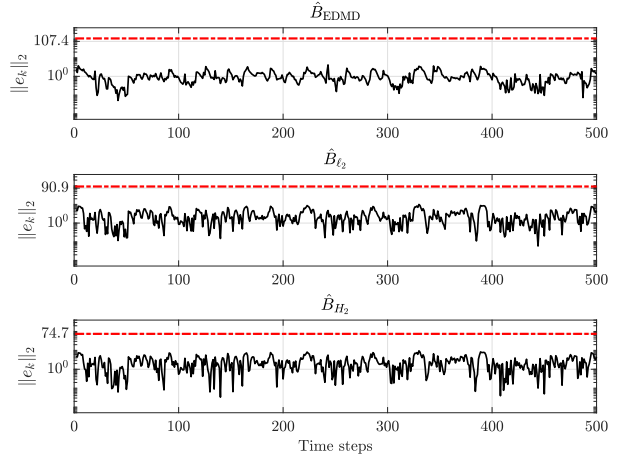}\vspace{-.4cm}    
\caption{Evolution of $\|e_k\|_2$ and error bound $\gamma_\text{amp}$ \eqref{eq:err_bound_thm} represented by (\protect\redline) for the approximated LTI models under white noise input.} 
\label{fig:err_bounds}
\end{center}
\end{figure}
In order to obtain the approximations $\hat{B}_{\ell_2}$ and $\hat{B}_{H_2}$ of $B_k$, we solve the minimization problems \eqref{eq:min_pb_1} and \eqref{eq:min_pb_2}, respectively. We do this by gridding the state and input space as follows: $x_{k,1}\in [-2.5,2.5]$ with a step of $0.05$, $x_{k,2}\in [-10,2.7]$ with a step of $0.25$ and $u_k\in [-1.6,2.1]$ with a step of $0.2$. Using this fine gridding results in more than 97000 points, which corresponds to the number of LMIs that need to be solved. To reduce the computational cost, one can use a random selection. In this example, it was observed that at 7000 grid points there was no drop in performance and the optimization problem was solved in a matter of minutes. The implementation was carried out in {\tt Matlab}, using the {\tt yalmip} toolbox \citep{Lofberg:04}. The resulted $\hat{B}_{\ell_2}$ and $\hat{B}_{H_2}$ with the respective $\gamma$-bounds are given below:
\begin{align*} 
\hat{B}_{\ell_2}=[1\; 3.3700\; -1.0600]^\top, \quad &\gamma_{\ell_2}=22.8026,\\
\hat{B}_{H_2}=[1\; 3.9602\; -0.2157]^\top, \quad &\gamma_{H_2}=9.1552.
\end{align*}
To obtain the most favourable $\hat{B}_{\text{EDMD}}$ to be used for comparison with the previously discussed approaches, we use the grid points in $(\mathbb{X},\mathbb{U})$ that correspond to the simulation trajectory of \eqref{eq:koop_model_example} in Section \ref{sec:simulation}. The data is collected as: $Z=[\Phi(x_0)\cdots\Phi(x_{N-1})]$, $Z^+=[\Phi(x_1)\cdots\Phi(x_{N})]$ and $U=[u_0\cdots u_{N-1}]$. Thus, as $A$ is known, the input matrix $\hat{B}_{\text{EDMD}}$ is computed as:
\begin{equation}
\hat{B}_{\text{EDMD}}=(Z^+-AZ)U^\dagger = [\ 1\;\ 0.4902\;\ 0.3093\ ]^\top.
\end{equation}
\vspace{-.5cm}
\subsection{Discussion and computation of bounds}
\vspace{-.2cm}
Table \ref{tb:gamma_values} shows the various error measures obtained via synthesis and analysis based on the $\ell_2$-gain and generalized $H_2$ norm. As expected, the lowest bounds $\gamma_{\ell_2}$ and $\gamma_{H_2}$ are the optimum values obtained in the synthesis procedures \eqref{eq:min_pb_1} and \eqref{eq:min_pb_2}. Solving the analysis problem, it can be observed that for both performance measures, the EDMD approximation produces higher bounds. Furthermore, we note that the optimal $\hat{B}_{H_2}$ produces the lowest error bound $\gamma_{\text{amp}}$, followed by $\hat{B}_{\ell_2}$ and $\hat{B}_{\text{EDMD}}$. The evolution of the 2-norm of the state error $e_k$ in \eqref{eq:error_in_z} is given in Fig. \ref{fig:err_bounds} for the introduced approximation methods. It can be seen that all trajectories satisfy the computed error bounds $\gamma_{\text{amp}}$.
\begin{figure}[t!]
\begin{center}
\includegraphics[width=.45\textwidth]{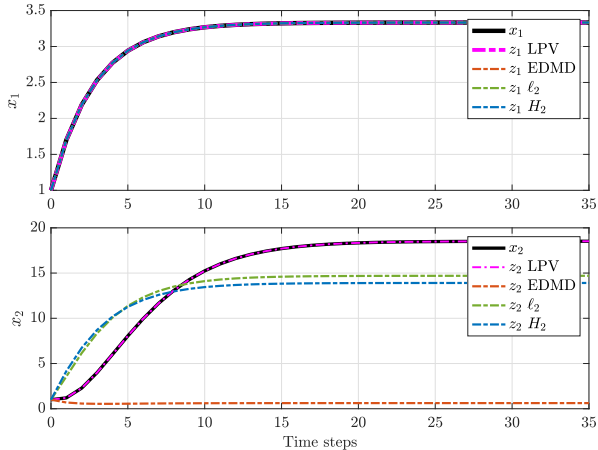}\vspace{-.4cm}    
\caption{Comparison between state trajectories of the nonlinear system \eqref{eq:nl_dt_ex}, the LPV Koopman model \eqref{eq:koop_model_example} and the LTI approximations under constant input.\vspace{-.1cm}} 
\label{fig:plot_constant}
\end{center}
\end{figure}
\par
\begin{figure}[t!]
\begin{center}
\includegraphics[width=.45\textwidth]{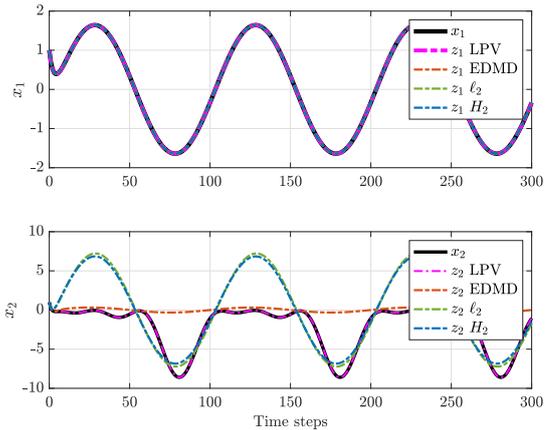}\vspace{-.4cm}    
\caption{Comparison between state trajectories of the nonlinear system \eqref{eq:nl_dt_ex}, the LPV Koopman model \eqref{eq:koop_model_example} and the LTI approximations under a sinusoidal input. \vspace{-.1cm}} 
\label{fig:plot_sinusoid}
\end{center}
\end{figure}

 We next analyze the behaviour of the approximated LTI Koopman models under both constant $u=1$ and varying $u=0.5\sin(2\pi t)$ inputs. As can be seen in Fig. \ref{fig:plot_constant}, both the $\ell_2$-gain and generalized $H_2$ norm-based models outperform the EDMD approximation for the constant input case. When the sinusoidal input is applied, as shown in Fig. \ref{fig:plot_sinusoid}, the first state trajectory is correctly characterized by all LTI approximations. For the second state, $x_2$, the EDMD model does not follow the dynamics of the original system, whereas the $\ell_2$-gain and generalized $H_2$ approximations follow the dynamics in the negative amplitude region. Overall, the synthesized models outperform the EDMD approximation and the provided analysis tools can efficiently characterise the expected performance of various approximation schemes to obtain LTI Koopman forms.\vspace{-.1cm}

\begin{table}[h!]
\begin{center}
\captionsetup{width=.48\textwidth}
\caption{Comparison of the approximation error (in terms of $\ell_2$-gain, generalised $H_2$ norm and $\gamma_\text{amp}$) of LTI Koopman models obtained via the $\ell_2$-gain optimal, generalised $H_2$ norm optimal, and EDMD approaches.  
}\label{tb:gamma_values}
\begin{tabular}{|c||c|c|c|}
\hline
 \rule{0pt}{3ex}  & $\gamma_{\ell_2}$  & $\gamma_{H_2}$ & $\gamma_\text{amp}$\\[1ex] \hline\hline\hline
 $\ell_2 \text{ approx}$ & $22.8026$  & $9.4207
$ & $90.86$\\\hline
  $H_2 \text{ approx}$ & $23.5944$  & $9.1552$ & $74.65$\\\hline
   $\text{EDMD}$ & $36.8768$  & $14.2335$ & $107.39$\\\hline
\end{tabular}
\end{center}
\end{table}
\vspace{-.2cm}
\section{Conclusion}\label{sec:conclusion}
\vspace{-.3cm}
Exact Koopman modelling for nonlinear systems with inputs gives an LPV form. To approximate the lifted representation with fully LTI Koopman models, the present paper derives $\ell_2$-gain and generalized $H_2$ norm optimal schemes. The resulted models are shown to outperform the popular EDMD-based scheme in the Koopman literature while also producing lower error bounds. Future research will focus on investigating how to merge results from the LPV framework with Koopman models to fully exploit the benefits of this modelling framework.
\vspace{-.2cm}
\bibliography{references_lpvs}             

\begin{thebibliography}{18}
\providecommand{\natexlab}[1]{#1}
\providecommand{\url}[1]{\texttt{#1}}
\providecommand{\urlprefix}{URL }
\expandafter\ifx\csname urlstyle\endcsname\relax
  \providecommand{\doi}[1]{doi:\discretionary{}{}{}#1}\else
  \providecommand{\doi}{doi:\discretionary{}{}{}\begingroup \urlstyle{rm}\Url}\fi

\bibitem[{Bevanda et~al.(2021)Bevanda, Sosnowski, and Hirche}]{Bevanda:21}
Bevanda, P., Sosnowski, S., and Hirche, S. (2021).
\newblock Koopman operator dynamical models: Learning, analysis and control.
\newblock \emph{Annual Reviews in Control}, 197--212.

\bibitem[{Brunton et~al.(2021)Brunton, Budišić, Kaiser, and {Nathan Kutz}}]{Brunton:21}
Brunton, S.L., Budišić, M., Kaiser, E., and {Nathan Kutz}, J. (2021).
\newblock Modern {Koopman} theory for dynamical systems.
\newblock \emph{ArXiv}, Abs/2102.12086.

\bibitem[{Byrnes and Lin(1994)}]{Byrnes:94}
Byrnes, C. and Lin, W. (1994).
\newblock Losslessness, feedback equivalence, and the global stabilization of discrete-time nonlinear systems.
\newblock \emph{IEEE Transactions on Automatic Control}, 83--98.

\bibitem[{Caigny et~al.(2013)Caigny, Camino, Oliveira, Peres, and Swevers}]{Caigny:13}
Caigny, J.D., Camino, J.F., Oliveira, R.C.L.F., Peres, P.L.D., and Swevers, J. (2013).
\newblock Gain-scheduled dynamic output feedback control for discrete-time {LPV} systems.
\newblock \emph{Int. Journal of Robust and Nonlinear Control}, 535--558.

\bibitem[{Iacob et~al.(2022)Iacob, Tóth, and Schoukens}]{Iacob:22}
Iacob, L.C., Tóth, R., and Schoukens, M. (2022).
\newblock Koopman form of nonlinear systems with inputs.
\newblock \emph{ArXiv}, Abs/2207.12132.

\bibitem[{Kaiser et~al.(2021)Kaiser, {Nathan Kutz}, and Brunton}]{Kaiser:21}
Kaiser, E., {Nathan Kutz}, J., and Brunton, S.L. (2021).
\newblock Data-driven discovery of {Koopman} eigenfunctions for control.
\newblock \emph{Machine Learning: Science and Technology}, 035023.

\bibitem[{Koelewijn and Tóth(2021)}]{Koelewijn:21}
Koelewijn, P.J.W. and Tóth, R. (2021).
\newblock Incremental stability and performance analysis of discrete-time nonlinear systems using the {LPV} framework.
\newblock \emph{4th IFAC Workshop on LPV Systems}, 75--82.

\bibitem[{Korda and Mezić(2018)}]{Korda:18}
Korda, M. and Mezić, I. (2018).
\newblock Linear predictors for nonlinear dynamical systems: {Koopman} operator meets model predictive control.
\newblock \emph{Automatica}, 149--160.

\bibitem[{Löfberg(2004)}]{Lofberg:04}
Löfberg, J. (2004).
\newblock {YALMIP} : A toolbox for modeling and optimization in {MATLAB}.
\newblock \emph{In Proc. of the 2004 IEEE Int. Conf. on Robotics and Automation}, 284--289.

\bibitem[{Mauroy et~al.(2020)Mauroy, Mezić, and Susuki}]{Mauroy:20}
Mauroy, A., Mezić, I., and Susuki, Y. (eds.) (2020).
\newblock \emph{The {Koopman} Operator in Systems and Control: Concepts, Methodologies and Applications}.
\newblock Springer.

\bibitem[{Mohammadpour and Scherer(2012)}]{Mohammadpour:12}
Mohammadpour, J. and Scherer, C.W. (eds.) (2012).
\newblock \emph{Control of Linear Parameter Varying Systems with Applications}.
\newblock Springer.

\bibitem[{Ping et~al.(2021)Ping, Yin, Li, Liu, and Yang}]{Ping:21}
Ping, Z., Yin, Z., Li, X., Liu, Y., and Yang, T. (2021).
\newblock Deep {Koopman} model predictive control for enhancing transient stability in power grids.
\newblock \emph{Int. Journal of Robust and Nonlinear Control}, 1964--1978.

\bibitem[{Proctor et~al.(2016)Proctor, Brunton, and {Nathan Kutz}}]{Proctor:16}
Proctor, J.L., Brunton, S.L., and {Nathan Kutz}, J. (2016).
\newblock Dynamic mode decomposition with control.
\newblock \emph{SIAM Journal on Applied Dynamical Systems}, 142--161.

\bibitem[{Scherer(2001)}]{Scherer:01}
Scherer, C.W. (2001).
\newblock {LPV} control and full block multipliers.
\newblock \emph{Automatica}, 361--375.

\bibitem[{Scherer and Weiland(2015)}]{Scherer:15}
Scherer, C.W. and Weiland, S. (2015).
\newblock Linear matrix inequalities in control.
\newblock Lecture notes, Univ. Stuttgart.

\bibitem[{Surana(2016)}]{Surana:16}
Surana, A. (2016).
\newblock Koopman operator based observer synthesis for control-affine nonlinear systems.
\newblock \emph{In Proc. of the 55th Conf. on Decision and Control}, 6492--6499.

\bibitem[{Verhoek et~al.(2021)Verhoek, Koelewijn, Tóth, and Haesaert}]{Verhoek:21}
Verhoek, C., Koelewijn, P.J.W., Tóth, R., and Haesaert, S. (2021).
\newblock Convex incremental dissipativity analysis of nonlinear systems.
\newblock \emph{ArXiv}, Abs/2006.14201.

\bibitem[{Williams et~al.(2015)Williams, Kevrekidis, and Rowley}]{Williams:16}
Williams, M.O., Kevrekidis, I.G., and Rowley, C.W. (2015).
\newblock A data–driven approximation of the {Koopman} operator: Extending dynamic mode decomposition.
\newblock \emph{Journal of Nonlinear Science}, 1307--1346.

\end{thebibliography}








\end{document}